\documentclass[a4paper,twocolumn,english,english,aps]{revtex4}
\usepackage[T1]{fontenc}
\usepackage[latin1]{inputenc}
\usepackage{babel}
\usepackage{graphics}

\makeatletter

\providecommand{\LyX}{L\kern-.1667em\lower.25em\hbox{Y}\kern-.125emX\@}
\let\SF@@footnote\footnote
\def\footnote{\ifx\protect\@typeset@protect
    \expandafter\SF@@footnote
  \else
    \expandafter\SF@gobble@opt
  \fi
}
\expandafter\def\csname SF@gobble@opt \endcsname{\@ifnextchar[%]
  \SF@gobble@twobracket
  \@gobble
}
\edef\SF@gobble@opt{\noexpand\protect
  \expandafter\noexpand\csname SF@gobble@opt \endcsname}
\def\SF@gobble@twobracket[#1]#2{}

\usepackage{amssymb}

\input epsf
\makeatother
\makeatother

\makeatother
\begin{document}

\title{Stationary dark energy with a baryon-dominated era: solving the coincidence
problem with a linear coupling}

\author{Domenico Tocchini-Valentini and Luca Amendola}

\affiliation{Osservatorio Astronomico di Roma, Viale del Parco Mellini 84,  00136 Roma, Italy,
email: amendola@coma.mporzio.astro.it, tocchini@oarhp1.rm.astro.it}

\date{\today{}}

\begin{abstract}
We show that all cosmological models with an accelerated stationary
global attractor reduce asymptotically to a dark energy field with
an exponential potential coupled linearly to a perfect fluid dark
matter. In such models the abundance of the dark components reaches
a stationary value and therefore the problem of their present coincidence
is solved. The requirement of a vanishing coupling of the baryons
in order to pass local gravity experiments induces the existence of
an intermediate baryon-dominated era. We discuss in detail the properties
of these models and show that to accomodate standard nucleosynthesis
they cannot produce a microwave background consistent with observations.
We conclude that, among stationary models, only a time-dependent coupling
or equation of state might provide a realistic cosmology. 
\end{abstract}
\maketitle

\section{Introduction}

The recent observations of an accelerating expansion \cite{rie},
together with new cosmic microwave, large-scale structure and lensing
data, give a strong indication that the universe fluid is composed
of at least four different components: relativistic matter (\( \leq  \)0.01\%),
baryons (few per cent), dark matter (\( \sim  \)30\%) and dark energy
(\( \sim  \)70\%). The present amount of these components raises
some deep questions: Why are the dark matter and dark energy, which
supposedly have a different scaling with time, almost equal right
now \cite{zla}? Why are the baryons strongly suppressed with respect
to dark matter? Why is the coincidence between the dark components
relatively close to the equivalence between matter and radiation (as
pointed out by \cite{ark})? 

The problem of the present coincidence between dark energy and dark
matter could be simply solved if the two fluids have the same scaling
with time (let us call a system of fluids \( \rho _{1},\rho _{2} \)
with identical scaling a {}``stationary{}'' model, since \( d(\rho _{1}/\rho _{2})/dt=0 \)).
It is interesting to remark that such a scaling can soon be observationally
tested \cite{dalal}. It is well known that, assuming an exponential
potential for the scalar field representing the dark energy, the field
scales as the dominant component \cite{ratra, fer, liddle}, but this
does not produce acceleration. The question therefore arises of which
is the most general system of dark matter (modeled as a perfect fluid)
and dark energy (a scalar field) that allows an \emph{accelerated
stationary global attractor}. In this paper, following the arguments
of ref. \cite{pavzim}, we show that all models which contain an accelerated
stationary global attractor reduce asymptotically to a system characterized
by an exponential potential \emph{and} a linear coupling between the
two components. Systems of such a kind have been already analyzed
first in refs. \cite{wan, wet95} and successively by many authors
\cite{ame3, hol, bil, gasp, pav}, while their perturbations have
been studied in ref. \cite{ame2}. In order to pass local gravity
experiments \cite{dam}, we argue that it is necessary to break the
universality of the coupling, leaving the baryons uncoupled or weakly
coupled (see e.g. \cite{ame3, dam, cas, ame4, bm}). These assumptions
completely define our cosmology. 

The consequences of a linear coupling between the dark components
are manifold. First, as already remarked, this explains the cosmic
coincidence and the acceleration. Second, the accelerated regime explains
the decay of the baryons with respect to the dark components. Third,
the near coincidence of the equivalence between the luminous components
(baryons and radiation) and the beginning of the dark era (dark equivalence
from now) is automatically enforced. Fourth, the baryons are the dominant
component between the two equivalence times, producing a decelerated
epoch in which gravitational instability is effective. Fifth, the
present status of the universe is independent of the initial conditions,
being on a global attractor. In addition, it is to be noticed that
the model requires only constants of order unity in Planck units,
and that they are all fixed by the observations of the present energy
densities and the acceleration. 

Nothwithstanding these intriguing features, we show that the model
we present here fails in satisfying at the same time the nucleosynthesis
requirements together with producing an acceptable cosmic microwave
background (CMB) angular power spectrum. In fact if the conditions
for a standard nucleosynthesis are adopted, during the recent accelerated
regime a fast growth of the perturbations is induced, which in turn
causes an excessive integrated Sachs-Wolfe (ISW) effect, in contrast
with the observations. Nevertheless, we think that the dynamics we
discuss is interesting on its own. 

The conclusion we draw is that only models with a time-dependent coupling
or a potential more complicated that an exponential may contain an
accelerated global attractor and at the same time be compatible with
nucleosynthesis. A solution along these lines has been proposed in
\cite{amtoc} where a non-linear modulation of the coupling allows
nucleosynthesis to happen and structure to form in a regime of weak
coupling, while the acceleration is produced in the subsequent (present)
regime of strong coupling. In ref. \cite{ven} it has been shown that
a similar mechanism may be realized in superstring theories.

\section{Homogeneous solutions}

Let us show first why all two-fluid systems with a stationary accelerated
attractor reduce asymptotically to the one investigated below. Let
us write a generic coupled two-fluid systems with equations of state
\( p_{x}=(w_{x}-1)\rho _{x} \) in a flat Friedmann metric as \begin{eqnarray}
\dot{\rho }_{c}+3Hw_{c}\rho _{c} & = & \delta ,\\
\dot{\rho }_{\phi }+3Hw_{\phi }\rho _{\phi } & = & -\delta ,\label{cons} 
\end{eqnarray}
 where the subscript \( c \) stands for cold dark matter (CDM) and
the subscript \( \phi  \) for a scalar field. The Friedmann equation
is \[
3H^{2}=\kappa ^{2}\left( \rho _{\phi }+\rho _{c}\right) ,\]
 where \( \kappa ^{2}=8\pi  \) and \( G=c=1 \). As shown in ref.
\cite{pavzim}, the stationary condition \( d(\rho _{c}/\rho _{\phi })dt=0 \),
that is \( \rho _{\phi }=A\rho _{c} \), can be satisfied only if
\begin{equation}
\delta =\sqrt{3\rho _{c}}\kappa \left( w_{c}-w_{\phi }\right) \frac{A}{\sqrt{1+A}}\rho _{c}.
\end{equation}
 Putting \( w_{c}=1 \) and observing that \( A=\Omega _{\phi }/(1-\Omega _{\phi }) \),
we have \begin{equation}
\delta =\sqrt{\frac{3\kappa ^{2}\Omega _{\phi }}{2}}\frac{\eta -1}{\sqrt{1+\eta }}|\dot{\phi }|\rho _{c},
\end{equation}
 where \( \eta =2U/\dot{\phi }^{2} \), the ratio of the potential
to kinetic scalar field energy (notice that \( w_{\phi }=2/(1+\eta ) \)).
The stationary solution is accelerated if \[
\frac{\ddot{a}}{a}=\dot{H}+H^{2}=\kappa ^{2}(1+A)\rho _{c}\left( \frac{1-w_{\phi }}{2}\Omega _{\phi }-\frac{1}{6}\right) >0,\]
 which, for \( 2>w_{\phi }>0 \), can be realized only if \( \Omega _{\phi }>1/3 \)
(i.e. \( A>1/2 \)) and \begin{equation}
\eta >\frac{3\Omega _{\phi }+1}{3\Omega _{\phi }-1}>2.
\end{equation}

So far we repeated the steps of ref. \cite{pavzim}. Now, let us consider
the asymptotic behavior of \( \eta  \). If \( \eta \rightarrow 0 \),
the kinetic energy dominates over the potential energy, and the asymptotic
solution is not accelerated. If, on the other hand, \( \eta \rightarrow \infty  \),
the potential energy dominates, and the solution will be identical
to that of a cosmological constant. In fact, in this limit \( \dot{\phi }=0 \)
and \( w_{\phi }=0 \) and Eq. (\ref{cons}) gives \( \delta =0 \)
where \begin{equation}
\delta =\rho _{c}\sqrt{3\kappa ^{2}U(\phi )\Omega _{\phi }},
\end{equation}
 which implies \( \rho _{c}\rightarrow 0 \). Therefore, barring oscillatory
solutions, only if \( \eta \rightarrow const. \) the scalar field
behaves as stationary accelerated dark energy (clearly this case includes
also that of dark energy as a perfect fluid). When \( \eta  \) is
constant, the coupling reduces to \begin{equation}
\label{lincoup}
\delta =\sqrt{2/3}\kappa \beta |\dot{\phi }|\rho _{c},
\end{equation}
 with \begin{equation}
\label{etabeta}
\beta =\frac{3}{2}(\eta -1)\sqrt{\frac{\Omega _{\phi }}{1+\eta }},
\end{equation}
 which is the form we study below. Moreover, it is easy to show that
\( \eta = \)const. implies an exponential potential \( U=U_{0}e^{-\sqrt{2/3}\mu \kappa \phi } \)
where \begin{equation}
\label{mueta}
\mu =\frac{3}{\sqrt{\Omega _{\phi }(1+\eta )}}\left[ 1+\frac{1}{2}(\eta -1)(1-\Omega _{\phi })\right] .
\end{equation}
 The conclusion is that a linear coupling and an exponential potential
represents the only non-trivial asymptotic case of stationary accelerated
dark energy. It is not difficult to see that this theorem extends
also to a Brans-Dicke theory with an explicit coupling between matter
components. As will be shown below, such a solution is also a global
attractor in a certain region of the parameter space. 

The condition for acceleration, \( \eta >(3\Omega _{\phi }+1)/(3\Omega _{\phi }-1) \),
together with \( \Omega _{\phi }<1, \) imply in Eq. (\ref{etabeta})
that \begin{equation}
\beta >\sqrt{3}/2.
\end{equation}
 This limit is much larger than allowed by local gravity experiments
on baryons, which give at most \( \beta <0.01 \) (see e.g. \cite{dam, wet95}).
Therefore, the theory must break the universality of the coupling
and let the baryons be decoupled from dark energy. An immediate consequence
of the species-dependent coupling, so far unnoticed, can be seen by
observing that the energy density of the dark components scales as
\begin{equation}
\rho _{\phi }\sim \rho _{c}\sim a^{-3\frac{\mu }{\mu +\beta }}.
\end{equation}
 For any \( \beta >0 \) the energy density decays slower than in
the standard matter-dominated Friedmann universe. Therefore, any uncoupled
(or weakly coupled) component, as the baryons, decays faster than
the coupled ones (see also \cite{gasp}). 

The cosmology we study below is a more realistic version of the one
above: we include in fact radiation and baryons, both of which are
coupled to the dark components only through gravitation. Once baryons
and radiation decay away, we recover the stationary accelerated attractor.
The Einstein equations for our model have been already described in
\cite{ame3}, in which a similar model (but on a different attractor,
i.e., for different parameters) was studied (see also ref. \cite{nunes}).
Here we summarize their properties. The conservation equations for
the field \( \phi  \) , cold dark matter, baryons (\( b \) ), and
radiation (\( \gamma  \) ), plus the Friedmann equation, are \begin{eqnarray}
\ddot{\phi }+3H\dot{\phi }+U_{,\phi } & = & -\sqrt{2/3}\kappa \beta \rho _{c},\nonumber \\
\dot{\rho }_{c}+3H\rho _{c} & = & \sqrt{2/3}\kappa \beta \rho _{c}\dot{\phi },\nonumber \\
\dot{\rho }_{b}+3H\rho _{b} & = & 0,\label{sys} \\
\dot{\rho }_{\gamma }+4H\rho _{\gamma } & = & 0,\nonumber \\
3H^{2} & = & \kappa ^{2}\left( \rho _{c}+\rho _{b}+\rho _{\gamma }+\rho _{\phi }\right) ,\nonumber 
\end{eqnarray}
 where \( H=\dot{a}/a \) and \( U(\phi )=U_{0}e^{-\sqrt{2/3}\mu \kappa \phi } \)
(we put \( \dot{\phi } \) instead of \( |\dot{\phi }| \) for generality).
The coupling \( \beta  \) can be seen as the relative strength of
the dark matter-dark energy interaction with respect to the gravitational
force. The only parameters of our model are \( \beta  \) and \( \mu  \)
(the constant \( U_{0} \) can always be rescaled away by a redefinition
of \( \phi  \)). For \( \beta =\mu =0 \) we reduce to the standard
cosmological constant case, while for \( \beta =0 \) we recover the
Ferreira \& Joyce model of \cite{fer}. As shown in ref. \cite{ame1},
the coupling we assume here can be derived by a conformal transformation
of a Brans-Dicke model, which automatically leaves the radiation uncoupled.
To decouple the baryons one needs to consider a two-metric Brans-Dicke
Lagrangian as proposed in \cite{dam}. Additional theoretical motivations
for this kind of coupling have been put forward in ref. \cite{gasp}
and for coupled dark energy in general in ref. \cite{carr}. 

The system (\ref{sys}) is best studied in the new variables \cite{cop, ame3}
\( x=\frac{\kappa }{H}\frac{\dot{\phi }}{\sqrt{6}},\quad y=\frac{\kappa }{H}\sqrt{U/3},\quad z=\frac{\kappa }{H}\sqrt{\rho _{\gamma }/3} \)
and \( u=\frac{\kappa }{H}\sqrt{\rho _{b}/3} \) and the time variable
\( \alpha =\log a \). Then we obtain \begin{eqnarray}
x^{\prime } & = & \left( z^{\prime }/z-1\right) x-\mu y^{2}+\beta (1-x^{2}-y^{2}-z^{2}-u^{2}),\nonumber \\
y^{\prime } & = & \mu xy+y\left( 2+z^{\prime }/z\right) ,\nonumber \\
u^{\prime } & = & -3/2u+u\left( 2+z^{\prime }/z\right) ,\label{sys2} \\
z^{\prime } & = & -z\left( 1-3x^{2}+3y^{2}-z^{2}\right) /2,\nonumber 
\end{eqnarray}
 where the prime denotes derivation with respect to \( \alpha . \)
The CDM energy density parameter is obviously \( \Omega _{c}=1-x^{2}-y^{2}-z^{2}-u^{2} \)
while we also have \( \Omega _{\phi }=x^{2}+y^{2}, \) \( \Omega _{\gamma }=z^{2} \)
and \( \Omega _{b}=u^{2} \). The system is subject to the condition
\( x^{2}+y^{2}+z^{2}+u^{2}\leq 1 \). 

The critical points of system (\ref{sys2}) are listed in Tab. I,
where \( p \) is the scale factor exponent, \( a\sim \tau ^{p/1-p}=t^{p} \),
where \( g\equiv 4\beta ^{2}+4\beta \mu +18 \), and where we used
the subscripts \( b,c,r \) to denote the existence of baryons, matter
or radiation, respectively, beside dark energy. In Tab. II we report
the conditions of existence and stability of the critical points,
denoting \( \mu _{+}=(-\beta +\sqrt{18+\beta ^{2}})/2 \) and \( \mu _{0}=-\beta -\frac{9}{2\beta } \).

\begin{table*}

\caption{Critical points.}

\begin{tabular}{ccccccccc}
\hline 
Point &
 \( x \)&
 \( y \)&
 \( z \)&
 \( u \)&
 \( \Omega _{\phi } \)&
 \( p \)&
 \( w_{eff} \)&
 \( w_{\phi } \)\\
\hline
\( a \)&
 \( -\frac{\mu }{3} \)&
 \( \sqrt{1-\frac{\mu ^{2}}{9}} \)&
 \( 0 \)&
 0 &
 1 &
 \( \frac{3}{\mu ^{2}} \)&
 \( \frac{2\mu ^{2}}{9} \)&
 \( \frac{2\mu ^{2}}{9} \)\\
 \( b_{r} \)&
 \( -\frac{2}{\mu } \)&
 \( \frac{\sqrt{2}}{\left| \mu \right| } \)&
 \( \sqrt{1-\frac{6}{\mu ^{2}}} \)&
 0 &
 \( \frac{6}{\mu ^{2}} \)&
 \( \frac{1}{2} \)&
 \( \frac{4}{3} \)&
 \( \frac{4}{3} \)\\
 \( b_{c} \)&
 \( -\frac{3}{2\left( \mu +\beta \right) } \)&
 \( \frac{\sqrt{g-9}}{2\left| \mu +\beta \right| } \)&
 0 &
 \( 0 \)&
 \( \frac{g}{4\left( \beta +\mu \right) ^{2}} \)&
 \( \frac{2}{3}\left( 1+\frac{\beta }{\mu }\right)  \)&
 \( \frac{\mu }{\mu +\beta } \)&
 \( \frac{18}{g} \)\\
 \( b_{b} \)&
 \( -\frac{3}{2\mu } \)&
 \( \frac{3}{2\left| \mu \right| } \)&
 0 &
 \( \sqrt{1-\frac{9}{2\mu ^{2}}} \)&
 \( \frac{9}{2\mu ^{2}} \)&
 \( \frac{2}{3} \)&
 1 &
 1 \\
 \( c_{r} \)&
 0 &
 0 &
 1 &
 0 &
 0 &
 \( \frac{1}{2} \)&
 \( \frac{4}{3} \)&
 \( - \)\\
 \( c_{rc} \)&
 \( \frac{1}{2\beta } \)&
 0 &
 \( \sqrt{1-\frac{3}{4\beta ^{2}}} \)&
 0 &
 \( \frac{1}{4\beta ^{2}} \)&
 \( \frac{1}{2} \)&
 \( \frac{4}{3} \)&
 2 \\
 \( c_{c} \)&
 \( \frac{2}{3}\beta  \)&
 0 &
 0 &
 0 &
 \( \frac{4}{9}\beta ^{2} \)&
 \( \frac{6}{4\beta ^{2}+9} \)&
 \( 1+\frac{4\beta ^{2}}{9} \)&
 2 \\
 \( d \)&
 \( -1 \)&
 0 &
 0 &
 0 &
 1 &
 \( 1/3 \)&
 2 &
 2 \\
 \( e \)&
 \( +1 \)&
 0 &
 0 &
 0 &
 1 &
 \( 1/3 \)&
 2 &
 2 \\
 \( f_{b} \)&
 0 &
 0 &
 0 &
 1 &
 0 &
 \( 2/3 \)&
 1 &
 \( - \) \\
\hline
\end{tabular}
\end{table*}

In Fig. 1 we display the parameter space of the model, indicating
for any choice of the parameters which point is a global attractor
(notice that there is complete symmetry under \( \beta \rightarrow -\beta  \)
and \( \mu \rightarrow -\mu  \)). As in \cite{ame3}, in which the
baryons have been included only as a perturbation, there exists one
and only one global attractor for any choice of the parameters. The
explicit inclusion of the baryons induces here two new critical points
(\( b_{b} \) and \( f_{b} \) ); moreover, contrary to \cite{ame3},
all the critical points with non-vanishing radiation are always unstable. 

From now on, we focus our attention on those parameters for which
the global attractor is \( b_{c} \), the only critical point that
may be stationary and accelerated. In Fig. 1 we show as a grey region
the parameters for which this attractor is accelerated. When \( b_{c} \)
is the global attractor, the system goes through three phases: 

\emph{a}) the radiation dominated era (the saddle \( b_{r} \) ); 

\emph{b}) the baryon dominated era ( the saddle \( b_{b} \) ); 

\emph{c}) the dark energy era (the global attractor \( b_{c} \) ). 

The dynamics of the model is represented in Fig. 2 (trend of \( \Omega _{c,b,\gamma ,\phi } \))
and in Fig. 3 (\( w_{eff} \)). During the various phases, the scalar
field is always proportional to the dominant component, just as in
the uncoupled model of ref. \cite{fer}. The three eras are clearly
visible: first, the energy density is dominated by the radiation,
with a constant contribution from the scalar field and a vanishing
one from dark matter; then, the baryons overtake the radiation, and
the scalar field scale accordingly; finally, the system falls on the
final stationary accelerated attractor, where dark matter and dark
energy share the energy density and the baryons decay away. The two
parameters \( \beta  \) and \( \mu  \) are uniquely fixed by the
observed amount of \( \Omega _{c} \) and by the present acceleration
parameter (or equivalently by \( w_{eff} \) ). For instance, \( \Omega _{c0}=0.30 \)
and \( w_{eff}=0.33 \) gives \( \mu =8 \) and \( \beta =16 \),
values which have been used in Fig. 2 and Fig. 3. With this value
of \( \mu  \) we have during radiation \( \Omega _{\phi }=6/\mu ^{2}\simeq 0.09 \),
compatible with the nucleosynthesis constraints (see e.g. \cite{fer, fre}).
To be more conservative, values of \( \mu  \) bigger than \( 11.5 \)
would satisfy the requirement of having \( \Omega _{\phi }<0.045 \)
during nucleosynthesis as suggested in \cite{bhm} but the situation
would be qualitatively similar. Once \( \beta  \) , \( \mu  \) and
the present baryon and radiation abundances are fixed, the model is
completely determined, and the ratio of dark matter to dark energy
is independent of the initial conditions.

\begin{table}

\caption{Properties of the critical points.}

\begin{tabular}{cccc}
\hline 
Point &
 Existence &
 Stability &
 Acceleration \\
\hline
\( a \)&
 \( \mu <3 \)&
 \( \mu <\mu _{+},\mu <\frac{3}{\sqrt{2}} \)&
 \( \mu <\sqrt{3} \)\\
 \( b_{r} \)&
 \( \mu >\sqrt{6} \)&
 unstable \( \forall  \)\( \mu ,\beta  \)&
 never \\
 \( b_{c} \)&
 \( \left| \mu +\beta \right| >\frac{3}{2},\mu <\mu _{0} \)&
 \( \beta >0,\mu >\mu _{+} \)&
 \( \mu <2\beta  \)\\
 \( b_{b} \)&
 \( \mu >\frac{3}{\sqrt{2}} \)&
 \( \beta <0,\mu >\frac{3}{\sqrt{2}} \)&
 never \\
 \( c_{r} \)&
 \( \forall  \)\( \mu ,\beta  \)&
 unstable \( \forall  \)\( \mu ,\beta  \)&
 never \\
 \( c_{rc} \)&
 \( \left| \beta \right| >\frac{\sqrt{3}}{2} \)&
 unstable \( \forall  \)\( \mu ,\beta  \)&
 never \\
 \( c_{c} \)&
 \( \left| \beta \right| <\frac{3}{2} \)&
 unstable \( \forall  \)\( \mu ,\beta  \)&
 never \\
 \( d \)&
 \( \forall  \)\( \mu ,\beta  \)&
 unstable \( \forall  \)\( \mu ,\beta  \)&
 never\\
 \( e \)&
 \( \forall  \)\( \mu ,\beta  \)&
 unstable \( \forall  \)\( \mu ,\beta  \)&
 never\\
 \( f_{b} \)&
 \( \forall  \)\( \mu ,\beta  \)&
 unstable \( \forall  \)\( \mu ,\beta  \)&
 never  \\
\hline
\end{tabular}
\end{table}

The radiation equivalence occurs at a redshift \( z_{eq} \) given
by \( (1+z_{eq})=\Omega _{b0}/\Omega _{\gamma 0}\simeq 500 \) for
realistic values. The dark equivalence redshift \( z_{dark} \) can
be found equating the baryon density and the dark energy density.
From the conservation laws \begin{equation}
\rho _{B}\sim a^{-3},\text {{}}\rho _{C}\sim \rho _{\phi }\sim a^{-3\frac{\mu }{\mu +\beta }}\, ,
\end{equation}
 putting \( r=\beta /\mu  \) and approximating \( \Omega _{\phi 0}\simeq \frac{r}{1+r} \)
(valid for \( \beta ,\mu \gg 1 \) ) it turns out that \begin{equation}
\label{zdark}
1+z_{dark}=\left[ \frac{r}{\Omega _{b0}\left( 1+r\right) }\right] ^{\frac{1}{3}\left( 1+\frac{1}{r}\right) },
\end{equation}
 For \( r\simeq 2 \), we obtain \( z_{dark}\simeq 5 \). 

The three main observations we compare our model to, nucleosynthesis,
structure formation and present acceleration, are produced in turn
during the three eras. The background trajectory discussed above passes
the nucleosynthesis constraint, yields the observed acceleration and
explains the cosmic coincidence. However, as we show next, it gives
an exceedingly large integrated Sachs-Wolfe effect.
\begin{figure}
{\centering \resizebox*{9cm}{!}{\includegraphics{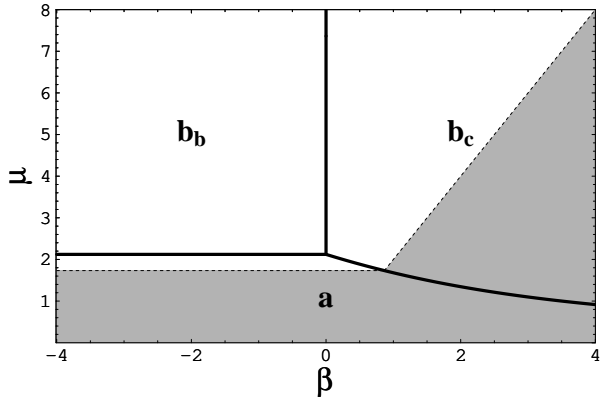}} \par}

\caption{Parameter space. Each region is labelled by the point that is a global
attractor there. Within the gray region the attractor is accelerated.}
\end{figure}
 
\begin{figure}
{\centering \resizebox*{9cm}{!}{\includegraphics{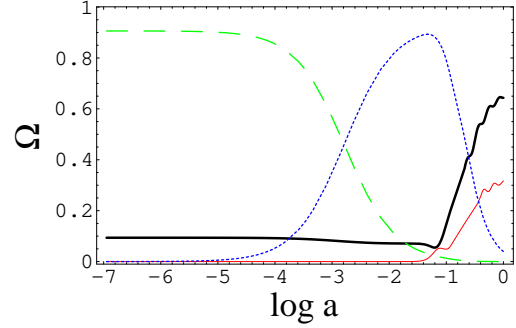}} \par}

\caption{Trend of the radiation (dashed green line), dark energy (thick black
line), dark matter (thin red line) and baryon (dotted blue line) density
fractions, for \protect\protect\( \mu =8\protect \protect \) and
\protect\protect\( \beta =16\protect \protect \) (here and in Figs.
3 and 6 the abscissa is \protect\protect\( \log _{10}a\protect \protect \).)}
\end{figure}

\begin{figure}
{\centering \resizebox*{9cm}{!}{\includegraphics{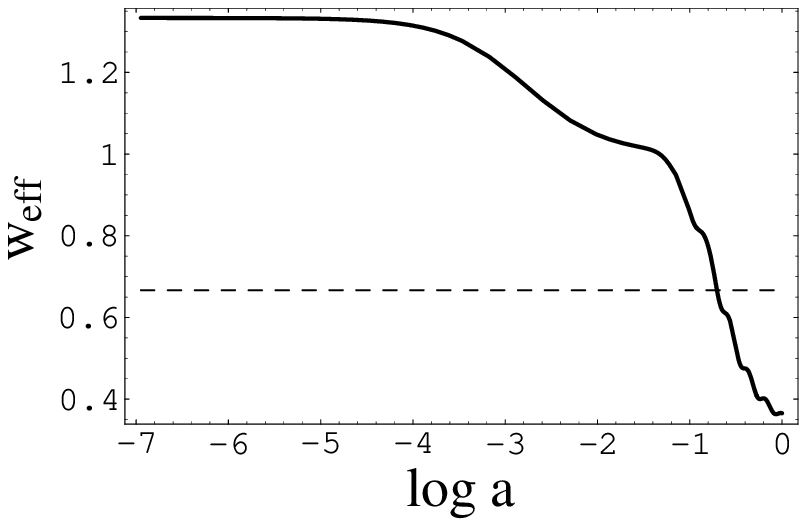}} \par}

\caption{The effective equation of state \protect\protect\( w_{eff}=1+p_{tot}/\rho _{tot}\protect \protect \)
for the same parameters as in the previous figure. Below the dashed
line the expansion is accelerated.}
\end{figure}

\section{Perturbations}

\noindent Close to the critical points \( b_{b} \) and \( b_{c} \)
the perturbations in the cold dark matter component (\( \delta _{c} \))
and in the baryonic one (\( \delta _{b} \)) grow as a power of the
scale factor, that is \( \delta _{c}=\delta _{b}/b=a^{m} \) (see
ref. \cite{bias})\( . \) In this expression the baryon bias \( b \)
and the growth exponent \( m \) depend only on the parameters \( \mu  \)
and \( \beta . \) Considering only the dominating growing modes,
during the baryonic phase the perturbations evolve asymptotically
with the law (see for instance ref. \cite{fer})\begin{equation}
m_{1}=\frac{1}{4}\left( -1+\sqrt{25-\frac{108}{\mu ^{2}}}\right) ;
\end{equation}
 while in the last plateau the common growth exponent is \[
m_{2}=\frac{1}{4\left( \beta +\mu \right) }\left[ -10\beta -\mu +\Delta \right] \]

\noindent where\[
\Delta ^{2}=-108+44\beta \mu +32\beta ^{3}\mu +25\mu ^{2}+\beta ^{2}\left( 32\mu ^{2}-44\right) \]
In ref. \cite{bias} it is also reported how restrictions on the baryon
bias would result in further constraints on the model but these will
not be used here since experimental bias determinations still remain
rather uncertain. The constraint provided by nucleosynthesis can surely
be considered on firmer grounds. 

The nucleosynthesis constraint \( \mu \gtrsim 7 \) (so that \( \Omega _{\phi }( \)1
MeV\( )\lesssim  \)\ \ 0.1) together with the limitation \( 0.6<\Omega _{\phi 0}<0.8 \)
implies a value of \( \beta  \) comprised between \( 9.8 \) and
\( 27.3 \). For this range of values the growth exponent in the last
era is found to lie between \( 7.4 \) and \( 15.3 \) respectively.
In Fig. 4 numerical evolutions of \( \delta _{c} \) and \( \delta _{b} \)
using the full set of equations are shown in the case \( \mu =7 \)
and \( \beta =9.8 \) for a fluctuation of wavelength \( 10 \) Mpc
\( h^{-1} \) : the fast growth during the final stage shows up clearly.
With such a conspicuous growth one expects a very large (late) ISW
effect on the CMB, which in fact appears in the numerical integrations
of the model of Fig. 5, produced using a version of CMBFAST modified
for the dark energy. The angular power spectrum is forced, by the
normalization procedure, to be highly suppressed at the lowest angular
scales respect to the observed values. This determines the failure
of the model investigated here. Even neglecting the nucleosynthesis
constraint, the existence of a radiation era requires \( \mu >\sqrt{6} \),
a value that induces again an unacceptably large ISW effect.

\begin{figure}
{\centering \resizebox*{10cm}{!}{\includegraphics{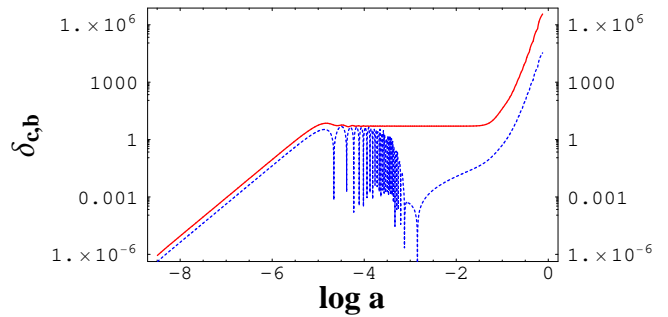}} \par}

\caption{The evolution of a 10 Mpc/\protect\( h\protect \) perturbation for
\(\protect\beta =9.8\) and \(\protect\mu =7\). The baryon fluctuations
\(\protect\delta _{b}\) are represented by the dotted (blue) line,
the CDM \(\protect\delta _{c}\) by the continuous (red) line.}
\end{figure}

\begin{figure}
{\centering \resizebox*{9cm}{!}{\includegraphics{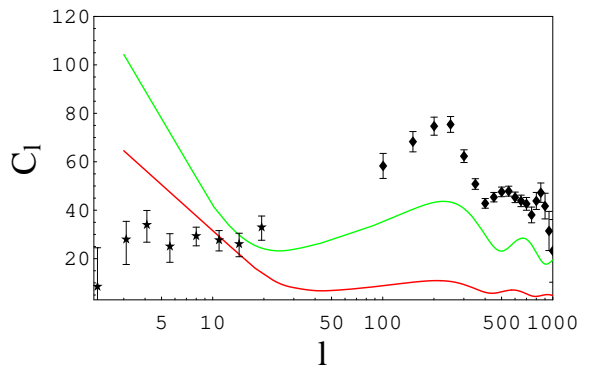}} \par}

\caption{CMB power spectra \protect\( C_{\ell }\protect \) for two set of
parameters: \protect\( \beta =4\protect \),\protect\( \mu =3.5\protect \)
(top curve) and \protect\( \beta =3.3\protect \),\protect\( \mu =4.2\protect \)
(bottom curve), compared with observational data from Boomerang \cite{net}
and COBE \cite{bon}. The other input values for CMBFAST are \protect\( h=0.8,\Omega _{b}=0.04,\Omega _{c}=0.3,n=1\protect \).
The strong ISW effect at small multipoles is evident. Larger values
of \protect\( \mu \protect \), as needed from the nucleosynthesis
constraint, enhance the problem.}
\end{figure}

\section{Conclusions}

Our coupled dark energy model provides a cosmological scenario with
quite unusual features. The standard sequence of a dark matter era
followed by a cosmological constant era, which forces to put the present
universe on a unlikely transient, is here replaced by a baryonic era
and a stationary dark era in which dark energy and dark matter share
a constant fraction of the total density. Contrary to almost all models
published so far, the present universe can be seen as already on the
final global attractor (except that, luckily, some baryons are still
around). The two new dimensionless constants introduced in our scenario,
\( \beta  \) and \( \mu  \), are determined by the present dark
matter energy density and by the present acceleration, and can be
of order unity. All cosmologies with an accelerated stationary global
attractor reduce asymptotically to the model discussed in this paper. 

In this model the coincidence problem is immediately solved by setting
\( \beta  \) and \( \mu  \) to the same order of magnitude. Regardless
of the initial condition, the universe evolves to a stationary state
with \( \Omega _{\phi }/\Omega _{c}= \)const. \emph{and} of order
unity. 

Moreover, this model explains also why the accelerated epoch occurs
\emph{just before} the present or, equivalently, why there are far
less baryons than CDM. The reason is provided by Eq. (\ref{zdark}):
fixing \( r=\beta /\mu  \) of order unity and \( \Omega _{b0} \)
of the order of a few per cent, we have \( z_{dark} \) near unity,
regardless of the initial conditions. That is, the fact that we observe
a relatively small quantity of baryons around implies that the accelerated
epoch is recent. Much more or much less baryons would push the beginning
of the accelerated epoch far in the future or in the past. 

The problem of the near coincidence between the radiation equivalence
and the dark equivalence can be rephrased as why \( z_{eq} \) and
\( z_{dark} \) are relatively close to each other. The answer is
that is the end of the radiation era that triggers the onset of the
baryon era, which in turn lasts for a relatively short time because
the system is heading toward the global attractor represented by the
dark era. 

Despite these positive features, the model as it stands cannot explain
our universe, since baryons and CDM fluctuations grow excessively
during the last accelerated phase if it is considered that a standard
nucleosynthesis has taken place earlier. This causes a CMB angular
power spectrum suppressed at the lowest angular scales respect to
the observational results. Thus we are forced to conclude that, for
the universe to fall on the stationary attractor, a non-linear modulation
in the coupling (as in ref. \cite{amtoc}), and/or a potential that
reduces only asymptotically to a pure exponential, is needed.

\end{document}